\UseRawInputEncoding
\RequirePackage[l2tabu, orthodox]{nag}
\documentclass[aps, prl, amsmath, twocolumn, longbibliography, superscriptaddress,floatfix, footinbib, 10pt, final]{revtex4-2}

\usepackage{hyphenat}
\usepackage{float}
\usepackage{graphicx} 
\usepackage{amssymb, amsmath}
\usepackage{blindtext}
\usepackage{siunitx} 
\usepackage[final]{microtype} 
\usepackage[T1]{fontenc}
\usepackage{lmodern}
\usepackage{xcolor}
\usepackage[colorlinks=true,linkcolor=blue, citecolor=blue, urlcolor=blue]{hyperref}
\usepackage[utf8]{inputenc}
\usepackage[acronym]{glossaries}
\usepackage{setspace}

\graphicspath{{./figures}}

\begin{document}

\newcommand{\kp}{\bm{k.p}}  
\newcommand{\RomanNumeralCaps}[1]
    {\MakeUppercase{\romannumeral #1}}

\newcommand{\polydept}{Department of Engineering Physics, \'Ecole Polytechnique de Montr\'eal, C.P. 6079, Succ. Centre-Ville, Montr\'eal, Qu\'ebec, Canada H3C 3A7}

\newcommand{\AppliedM}{Applied Materials Inc., 974 E. Arques Avenue, Sunnyvale, CA 94085, USA}

\newcommand{\SiGe}[2]{Si$_{#1}$Ge$_{#2}$}

\newcommand{\ac}[1]{\gls*{#1}}


\newcommand{\citesupp}{
    \footnote{See Supplemental Material at \textbf{[URL will be inserted by publisher]} for details of the $14-$band $k\cdot p$ theoretical formalism, as well as the SE, Raman, AFM, and APT experimental characterization, which includes Refs. \cite{Tsang1994, VandeWalle1989, Laude1971, Foreman1997, Djurisic2000, Palik1998, Ridene2001InfraredWells, Bahder1990, Weber1989, Ivchenko1996Heavy-lightIncidence,Foreman1998AnalyticalMixing, Fujiwara2007, McSkimin1953, Cukaric2013, Balslev1966, VandeWalle1986, Paul2016, Vina1984, DeSalvador2000, Herzinger1998, Varshni1967, Dismukes1964, Rideau2006, Elkurdi2003, Reeber1996, Tompkins2005, Chandrasekhar1977, Madelung2002, Qiao2012, Szmulowicz2004, Lautenschlager1987TemperatureSilicon}}\nocite{Tsang1994, VandeWalle1989, Laude1971, Foreman1997, Djurisic2000, Palik1998, Ridene2001InfraredWells, Bahder1990, Weber1989, Ivchenko1996Heavy-lightIncidence, Foreman1998AnalyticalMixing, Fujiwara2007, McSkimin1953, Cukaric2013, Balslev1966, VandeWalle1986, Paul2016, Vina1984, DeSalvador2000, Herzinger1998, Varshni1967, Dismukes1964, Rideau2006, Elkurdi2003, Reeber1996, Tompkins2005, Chandrasekhar1977, Madelung2002, Qiao2012, Szmulowicz2004, Lautenschlager1987TemperatureSilicon}
}


\newacronym{sls}{SLs}{superlattices}
\newacronym{cb}{CB}{conduction band}
\newacronym{sl}{SL}{superlattice}
\newacronym{if}{IF}{interface}
\newacronym{ml}{ML}{monolayers}
\newacronym{se}{SE}{spectroscopic ellipsometry}
\newacronym{qw}{QW}{quantum well}
\newacronym{apt}{APT}{atom probe tomography}
\newacronym{hrxrd}{HRXRD}{high-resolution Xray diffraction}
\newacronym{aoi}{AOI}{angle of incidence}
\newacronym{cp}{CP}{critical point}
\newacronym{rpcvd}{RPCVD}{reduced-pressure chemical vapor deposition}
\newacronym{haadf-stem}{HAADF-STEM}{high-angle annular dark field scanning transmission electron microscopy}
\newacronym{de}{DE}{differential evolution}
\newacronym{rms}{RMS}{surface roughness}
\newacronym{eels}{EELS}{electron energy loss spectroscopy}
\newacronym{rta}{RTA}{rapid thermal annealing}
\newacronym{afm}{AFM}{atomic force microscopy}
\newacronym{rtse}{RTSE}{room temperature spectroscopic ellipsometry}

\title{Localized Energy States Induced by Atomic-Level Interfacial Broadening in Heterostructures}

\author{Anis Attiaoui}
\thanks{These authors contributed equally to this work.}
\affiliation{\polydept{}}

\author{Gabriel Fettu}
\thanks{These authors contributed equally to this work.}
\affiliation{\polydept{}}

\author{Samik Mukherjee}
\affiliation{\polydept{}}

\author{Matthias Bauer}
\affiliation{\AppliedM{}}

\author{Oussama Moutanabbir}
\email{oussama.moutanabbir@polymtl.ca}
\affiliation{\polydept{}}

\begin{abstract}
A theoretical framework incorporating atomic-level interfacial details is derived to include the electronic structure of buried interfaces and describe the behavior of charge carriers in heterostructures in the presence of finite interfacial broadening. Applying this model to ultrathin heteroepitaxial (\SiGe{1-x}{x})$_m$/(Si)$_m$ superlattices predicts the existence of localized energy levels in the band structure induced by sub-nanometer broadening, which provides additional paths for hole-electron recombination. These predicted interfacial electronic transitions and the associated absorptive effects are confirmed experimentally at variable superlattice thickness and periodicity. By mapping the energy of the critical points, the optical transitions are identified between $2$ and $\SI{2.5}{\electronvolt}$ thus extending the optical absorption to lower energies. This phenomenon enables a straightforward and non-destructive probe of the atomic-level broadening in heterostructures.        
\end{abstract}
\maketitle

\UseRawInputEncoding
\par Interfaces are ubiquitous in design and processing of a variety of low-dimensional systems and devices \cite{Scappucci2021TheRoute,Paul2010TheSubstrates,Grange2020Atomic-ScaleScattering}. Their characteristic features such as strain field, composition, and topology are known to depart from the idealized picture of atomically flat and abrupt joint surfaces \cite{Grange2020Atomic-ScaleScattering,Leontiou2010SuppressionLayers,Luna2012CriticalHeterointerfaces,Moutanabbir2012DynamicGrowth}. The nature of these smeared interfaces impacts the behavior of charge carriers and shape the overall optical and electronic characteristics. These effects become even more prominent as the device dimension shrinks and the operating principles involve subtle quantum processes. For instance, the three-dimensional interfacial roughness has been shown to influence the confinement of charge carriers and their wavelength spread across one or more interfaces. As a matter of fact, this roughness must be considered to accurately describe the scattering mechanisms and predict the heterostructure optoelectronic properties \cite{Sakaki1987InterfaceWells,Califano2007InterwellRole,Grange2020Atomic-ScaleScattering}. This interplay between interfacial roughness and carrier scattering has also been critical in the design of highly scaled nanosheet transistors and spin qubits \cite{Jang2017DeviceNode,Wuetz2021AtomicDots}. In the latter, for instance, the atomic-level disorder at a quantum well interface was found to induce a large spread in valley splitting energy thereby hindering the uniformity of electron spin qubits \cite{Wuetz2021AtomicDots}.
\par
\begin{figure*}[htp]
    \centering
    \includegraphics[width=0.9\textwidth]{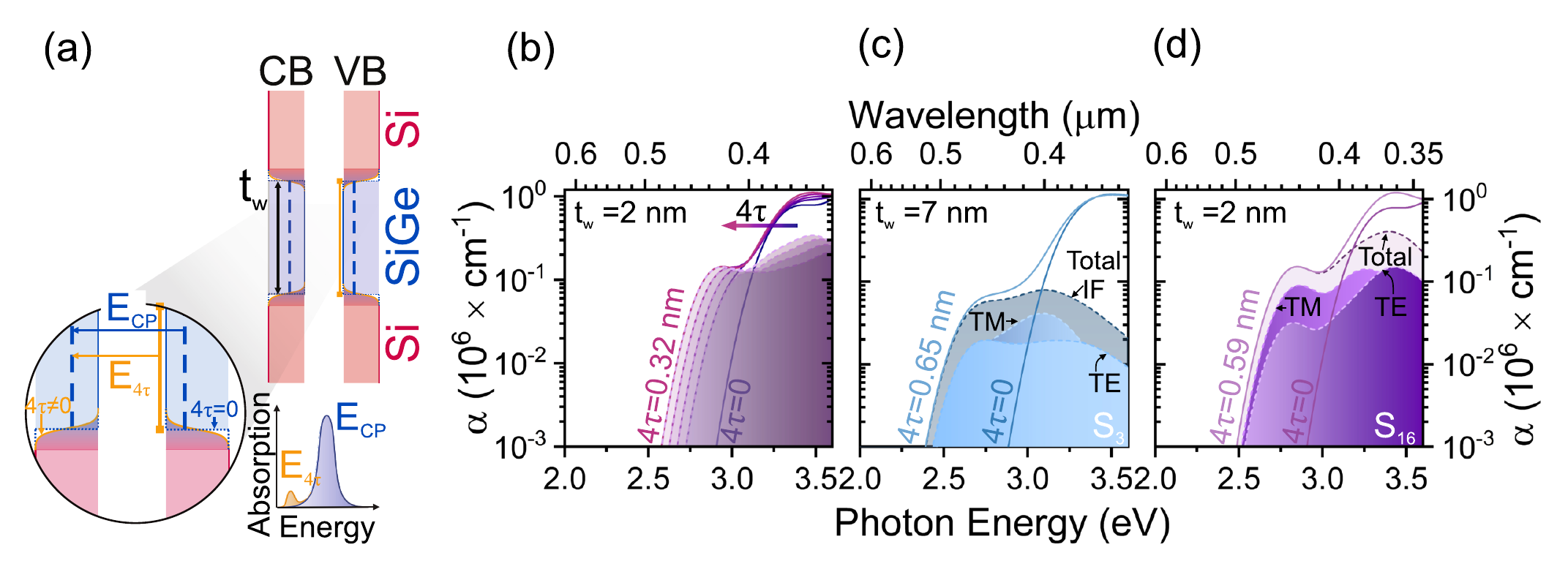}
    \caption{\textbf{The predicted effect of interfacial broadening on the optical absorption.} (a) Schematic representation of a \SiGe{}{}/Si \ac{qw} with (orange) and without (blue) the effect of interfaces broadening. The \ac{qw} has a thickness of $t_w$. The zoom-in inset highlights both possible optical transitions. (b) Next, the 14\hyp{}band $k\cdot p$ absorption coefficient $(\alpha)$ of the \ac{qw} with a variable interfacial broadening $4\tau$ from $0$ to $\SI{0.32}{\nano\meter}$ with a $\SI{0.1}{\nano\meter}$ step. TE and TM polarization-dependent absorption coefficient $(\alpha)$ of 2 \ac{sls}: (c) $S_3$ and (d) $S_{16}$ with APT-measured input parameters (thicknesses: $S_3\rightarrow t_w=7.3\pm \SI{0.2}{\nano\meter}, S_{16}\rightarrow t_w=2.2\pm \SI{0.3}{\nano\meter}$, interfacial width $(4\tau): S_3\rightarrow 0.65\pm\SI{0.10}{\nano\meter}, S_{16}\rightarrow 0.59\pm\SI{0.20}{\nano\meter}$, Ge content: $S_3\rightarrow 29.3\pm2.3\;\%, S_{16}\rightarrow 24.7\pm3.0\%)$.}
    \label{fig:fig1theory}
\end{figure*}

 It is clear that the behavior of charge carriers in heterostructures is shaped by the interfacial broadening, which is typically on the order of a few \ac{ml} \cite{Leontiou2010SuppressionLayers,Luna2012CriticalHeterointerfaces,Moutanabbir2012DynamicGrowth}. Herein, we argue that this atomic-level broadening creates localized energy levels yielding a distinct optical transition. Using \SiGe{}{}/Si as a model system, Figure \ref{fig:fig1theory}(a) illustrates the basic band-to-band absorption for a type$-$\RomanNumeralCaps{1} \SiGe{}{}/Si \ac{qw} along with the localized energy levels induced in the band structure at the interface. These levels provide additional paths for electron-hole recombinations, as highlighted by the direct transition, labeled hereafter $E$\textsubscript{4$\tau$} (Fig. \ref{fig:fig1theory}(a), inset). In principle, this transition would manifest as an additional absorption signature at lower energy when compared to the main interband \ac{cp} absorption peak $(E$\textsubscript{CP}$)$.

\UseRawInputEncoding
 \par To evaluate the hypothesized interface-related changes in the band structure, a framework was first implemented to quantitatively include the absorptive effects of buried interfaces. The miniband structure as well as the electron and hole wavefunctions are calculated within the 14\hyp{}band $k\cdot p$ formalism, where the microscopic effect of the interface is accounted for through the interface asymmetry Hamiltonian $(H$\textsubscript{IF}$)$ \cite{Ivchenko1996Heavy-lightIncidence,Foreman1998AnalyticalMixing,Szmulowicz1995DerivationApproximation}. Both  Si-on-\SiGe{}{} and \SiGe{}{}-on-Si interfaces are considered in $H$\textsubscript{IF}. Since in Si-rich \ac{sls} the lowest \ac{cb} has the $\Gamma_4^-/(\Gamma_8^-+\Gamma_6^-)$ symmetry and the $\Gamma_2^-$ \ac{cb} is close in energy, the four lowest \ac{cb}s were considered for an accurate simulation of the interband absorption \cite{Ridene2001InfraredWells}. To that end, a new \ac{cb} parametrization for Si and \SiGe{1-x}{x} alloys was developed within the 14\hyp{}band $k\cdot p$ theory; see \textcolor{blue}{Supplemental Material \citesupp{}}. The optical transition energies are calculated between electron and hole minibands based on crystal momentum conservation, and the electron-hole wavefunction overlap integral indicates which transitions are strongly active in absorption. The interface width $(4\tau)$ was varied between 0 and a few \ac{ml} to capture the broadening effects, where $4\tau$ = 0 corresponds to an abrupt interface.
\par Figure \ref{fig:fig1theory}(b) displays the calculated absorption coefficient $(\alpha$ in $cm^{-1})$ for a \SiGe{0.71}{0.29}/Si \ac{qw} at different values of $4\tau$ between 0 and $\SI{0.32}{\nano\meter}$. The well thickness $(t_w)$ was fixed at $\SI{2}{\nano\meter}$. Interestingly, non-zero interfacial widths were found to yield a distinct absorption peak below $\SI{3}{\electronvolt}$ whose energy redshifts as $4\tau$ increases. This behavior is confirmed by evaluating the interface contribution to $\alpha$ for two sets of (\SiGe{0.73}{0.27})$_m$/(Si)$_m$ \ac{sls} with a periodicity \textit{m} = 3 and 16 and a fixed total thickness of $\SI{60}{\nano\meter}$. These \ac{sls} are labeled hereafter as $S_m$. The well thickness in $S_3$ and $S_{16}$ is $\SI{7}{\nano\meter}$ and $\SI{2}{\nano\meter}$, respectively. The obtained polarized-dependent absorption coefficients are displayed in Fig. \ref{fig:fig1theory}(c, d), where $\alpha$\textsubscript{Total-IF} is defined as $\;2\alpha$\textsubscript{TE}$+\alpha$\textsubscript{TM}. Note that the stronger the interface potential, the greater is the interface state energy. Furthermore, the intensity of the interface\hyp{}related transitions is higher in $S_{16}$ \ac{sl} than $S_3$, which reflects the relative importance of the interface as $t_w$ of $S_{16}$ $(\sim \SI{2}{\nano\meter})$ is thinner than that of $S_3$ $(\sim \SI{7}{\nano\meter})$. Additionally, $\alpha$\textsubscript{TM} is broader in $S_3$ than $S_{16}$. This is directly associated to the number of available interface\hyp{}related transitions. In fact, there are fewer confined \ac{cb} states in $S_{16}$ than in $S_3$, which are converted to interface-promoted transitions, hence the sharper absorption coefficient. Therefore, the effect of interfaces is more prominent for thinner layers.   
\par To evaluate experimentally the theoretical findings, a series of ultra-short \SiGe{}{}/Si \ac{sls} were epitaxially grown at a Ge content below $30\%$ in a \ac{rpcvd} reactor.  Four (\SiGe{1-x}{x})$_m$/(Si)$_m$ \ac{sls} with different periodicity \textit\textit{m} = 3, 6, 12, or 16 were investigated. $S_{16}$ and $S_{12}$ were grown at $\SI{650}{\degree}$C, $S_6$ at $\SI{600}{\degree}$C, and $S_3$ at $\SI{500}{\degree}$C \cite{Mukherjee20203DSuperlattices}. 
\begin{figure*}[htp]
    \centering
    \includegraphics[width=.76\textwidth]{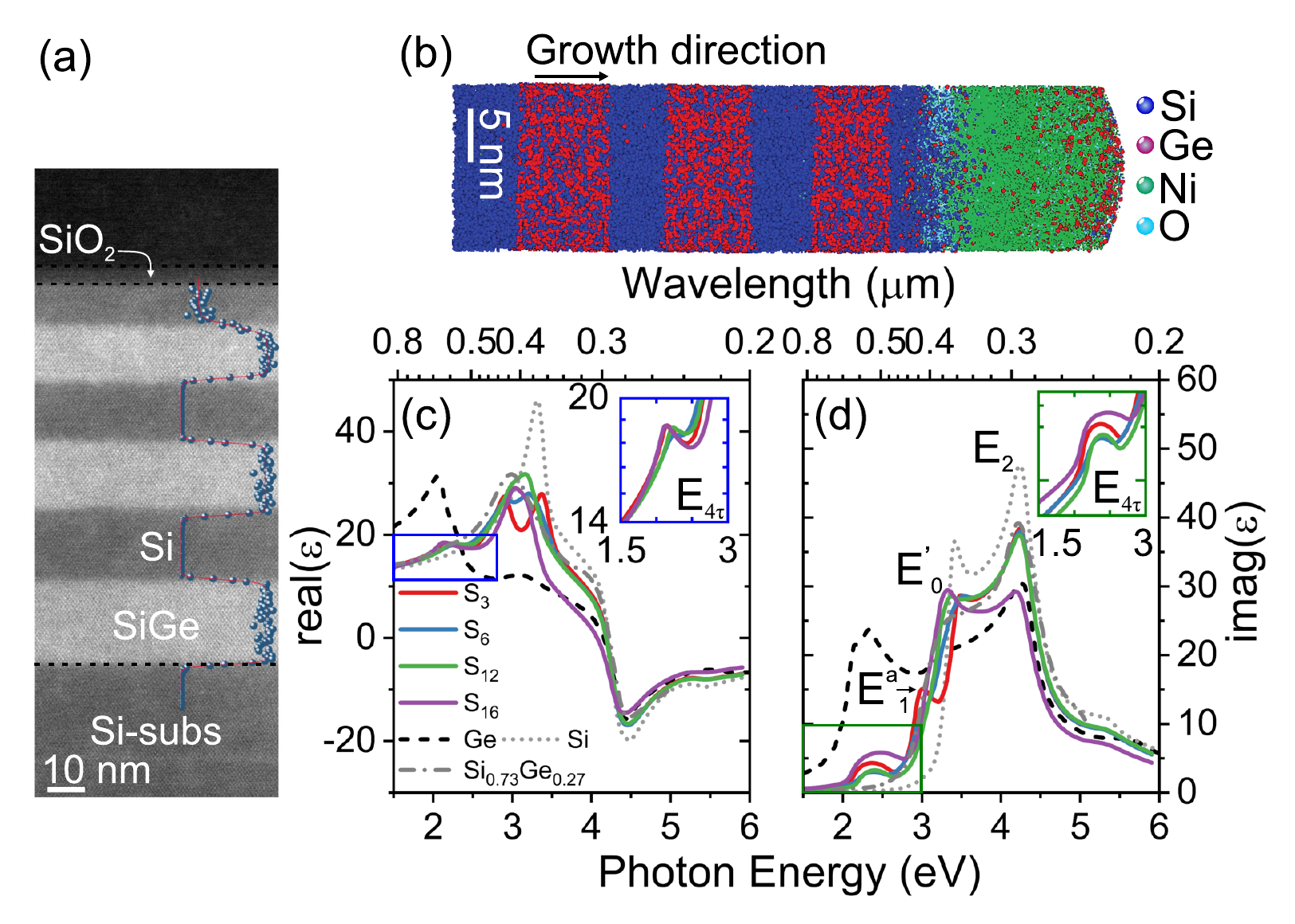}
    \caption{\textbf{Structural and optical characterization.} HAADF-STEM image of the (\SiGe{0.71}{0.29})$_3$/(Si)$_3$ SL overlaid with the concentration profile of Ge extracted from the corresponding \ac{apt} $3D$ reconstruction shown in panel (b). The average well and barrier thicknesses were estimated by \ac{apt} to be $7.3\pm\SI{0.2}{\nano\meter}$ and $6.0\pm\SI{0.2}{\nano\meter}$, respectively. (c) The real $(\varepsilon_1)$ and imaginary  $(\varepsilon_2)$ part of the complex dielectric function of the four (\SiGe{1-x}{x})$_m$/(Si)$_m$ \ac{sls} with increasing periodicity $\textit{m}(=3, 6, 12, 16)$ are presented. The Ge \cite{Aspnes1983}, Si \cite{Herzinger1998EllipsometricInvestigation} bulk and $\SI{90}{\nano\meter}$ thick \SiGe{0.73}{0.27} \cite{RajaMuthinti2012Effects001} dielectric functions are also shown for comparison purposes. The blue and green insets are log-log plot of $\varepsilon_1$ and $\varepsilon_2$, respectively between the energy range $(1.5 - 3\,eV)$. They highlight the interfacial broadening related optical transition $E$\textsubscript{4$\tau$}.}
    \label{fig:fig2MTR}
\end{figure*}
Figure \ref{fig:fig2MTR}(a) shows the \ac{haadf-stem} image of $S_3$ along with Ge concentration profile obtained using atom probe tomography (\ac{apt}). A representative $3D$ atom\hyp{}by\hyp{}atom \ac{apt} reconstruction map is also displayed in Fig. \ref{fig:fig2MTR}(b). \ac{apt} analysis was exploited to quantify the interfacial width $4\tau$ and the atomic content in each layer of the \ac{sls} by following the procedure described in Ref.\cite{Dyck2017AccurateTomography}. The \SiGe{}{}/Si stack is fully strained with an average degree of strain relaxation of $5.4\%$. Additionally, \ac{rms} was measured by \ac{afm}. Details of structural characterization are shown in \textcolor{blue}{Supplemental Material \cite{Note1}}. The interband optical absorption of the \ac{sls} was measured at room temperature by variable angle \ac{se}. The ellipsometric parameters $\Psi$ and $\Delta$ are acquired at an \ac{aoi} between $\SI{55}{\degree}$ and $\SI{80}{\degree}$ with a $\SI{5}{\degree}$ step covering photon energy $\omega$ from $1.5$ to $\SI{6}{\electronvolt}$. Since the \ac{sl} region has distinct optical properties than its constituent materials, the \ac{sl} is modeled as a single layer with its own unique set of optical constants \cite{Snyder1988MeasurementEllipsometry,Woollam1990EllipsometricStructures,Wagner1998SpectroscopicSuperlattices}.

\par The ``wavelength\hyp{}by\hyp{}wavelength'' energy\hyp{}dependent complex dielectric functions $(\tilde{\varepsilon}= \varepsilon_1+i\varepsilon_2)$ are shown for all \ac{sls} in Fig. \ref{fig:fig2MTR}; more details in \textcolor{blue}{Supplemental Material \cite{Note1}}. Moreover, overlaid in Fig. \ref{fig:fig2MTR}(c,d) are the data for bulk Ge \cite{Aspnes1983} (dashed\hyp{}black), bulk Si \cite{Herzinger1998EllipsometricInvestigation} (gray-dotted line) and fully strained, $\SI{90}{\nano\meter}$ thick \SiGe{0.73}{0.27} \cite{RajaMuthinti2012Effects001} dielectric functions (gray dash\hyp{}dotted). The intensity range of the optical properties of all \ac{sls} as well as the main spectral \ac{cp} $(E_2, E_0^\prime)$ positions qualitatively agree with those of bulk materials. The \ac{sl}-embedded \SiGe{1-x}{x} layers play an important role in modulating $\tilde{\varepsilon}$ of the whole SL given that its intensity is between that of bulk Si and bulk Ge. Furthermore, the $E_1$ band edge for the \SiGe{0.73}{0.27} dielectric function has an onset of $\SI{2.92}{\electronvolt}$ to reach an inflection point near $\SI{3.2}{\electronvolt}$. For a Ge content below $30\%$, the $E_1$ \ac{cp} is a superposition between the Si $E_0^\prime$ \ac{cp} located at $\SI{3.35}{\electronvolt}$ \cite{Lautenschlager1987TemperatureSilicon} and the \SiGe{}{} $E_1$ \ac{cp}. 

\par
\begin{figure}[t]
    \centering
    \includegraphics[width=.75\columnwidth]{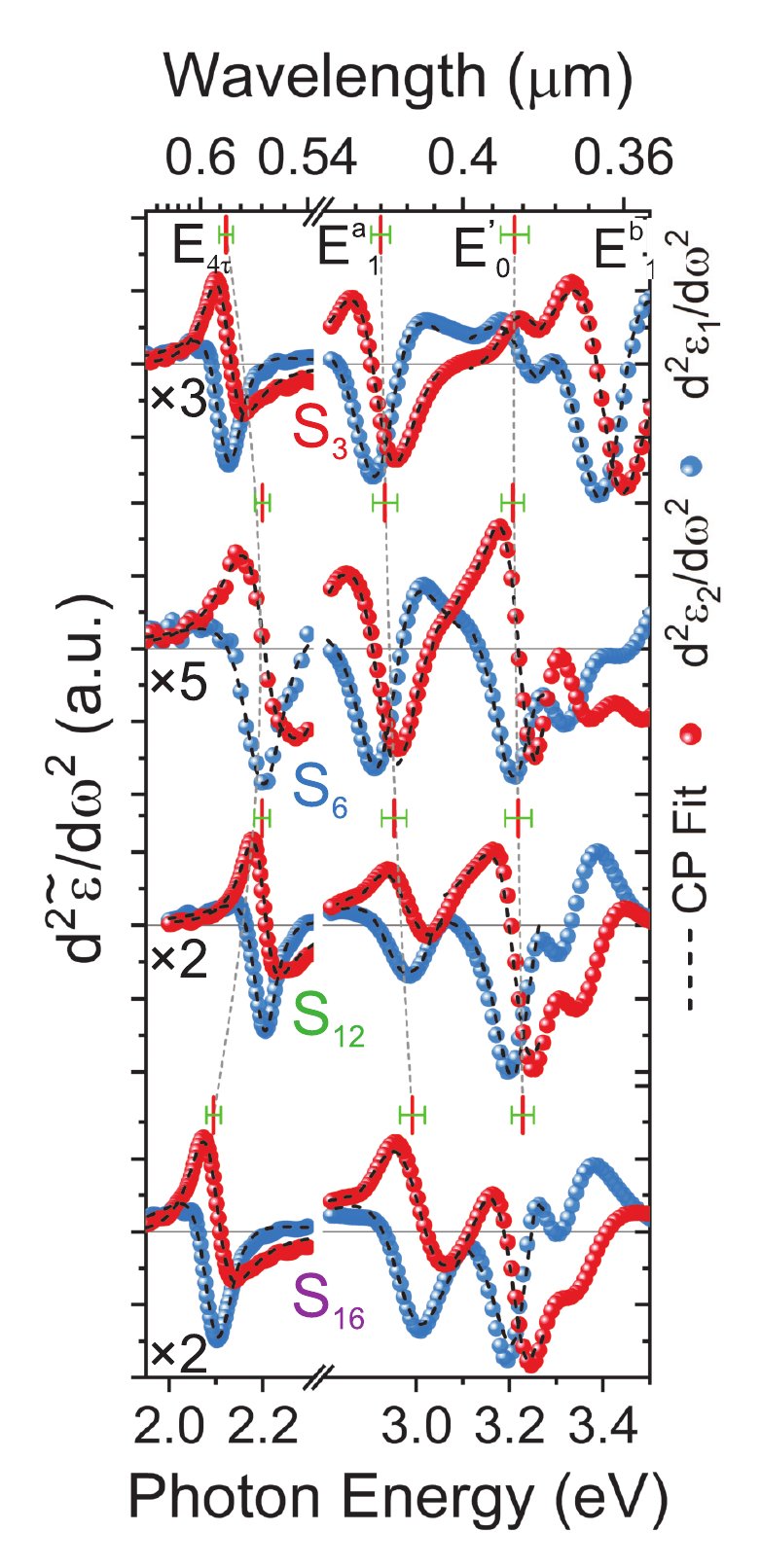}
    \caption{\textbf{Second derivative \ac{cp} analysis.} The second derivative of $\tilde{\varepsilon}$ as well as the \ac{cp} lineshape fit (dashed-black lines). The energy error bars are a combination between the SE experimental error $(\sim \SI{5}{\milli\electronvolt})$ and the 95\% confidence fit error. The dashed\hyp{}gray lines are a guide to the eye for the evolution of the \ac{cp}s as a function of the \ac{sl} periodicity \textit{m}. The vertical red lines are the peak position for each \ac{cp}, evaluated from the fit. Note that the vertical scale has to be divided by the factor given under each spectrum.}
    \label{fig:fig3d2e2}
\end{figure}

 The dielectric functions of all \ac{sls} clearly indicate the presence of an additional broad low-intensity peak below $\SI{3}{\electronvolt}$ that is absent in both Si bulk and \SiGe{}{} thin film. Only Ge bulk exhibits a strong \ac{cp} near $\SI{2.1}{\electronvolt}$ related to the $E_1$ transition occurring along the eight equivalent $[111]$ directions of the Brillouin zone (dashed-black line in Fig. \ref{fig:fig2MTR}(c)). The observed \ac{sl}\hyp{}related \ac{cp} peak position between $2$ and $\SI{2.5}{\electronvolt}$ agrees well with the theoretical predictions highlighted in Fig. \ref{fig:fig1theory}(b-d). Thus, the origin of this \ac{cp}, henceforth labeled as $E$\textsubscript{4$\tau$}, will be discussed in light of the interfacial broadening. The blue and green insets in Fig. \ref{fig:fig2MTR}(c, d) are zoom\hyp{}in log\hyp{}log plots, between $\SI{1.5}{\electronvolt}$ and $\SI{3}{\electronvolt}$, confirming the overlap between the CP energy and the calculated $E$\textsubscript{4$\tau$} transition. The $E$\textsubscript{4$\tau$}, $E_0^\prime$ and $E_1$ \ac{cp}s in Fig. \ref{fig:fig2MTR}(c, d) are analyzed in Fig. \ref{fig:fig3d2e2}. The second order dielectric function derivatives as well as the corresponding fits for the \ac{cp}s are also shown for all the \ac{sls}. The generic standard critical\hyp{}point lineshape model \cite{Aspnes1982OpticalFilms,Lautenschlager1987TemperatureSilicon} was used to fit the second derivative of $\tilde{\varepsilon}$. The real and imaginary parts of $(d^2\tilde{\varepsilon}/d\omega^2)$ were fitted simultaneously using a global optimization procedure based on the \ac{de} algorithm, as specified in \textcolor{blue}{Supplemental Material \cite{Note1}}. Additionally, the obtained \ac{cp} energies are shown for each \ac{sl} as a vertical red line.
\par The $E$\textsubscript{4$\tau$} transition is a distinct peak in $d^2\tilde{\varepsilon}/d\omega^2$ between $\SI{2}{\electronvolt}$ and $\SI{2.5}{\electronvolt}$ for all the (\SiGe{1-x}{x})$_m$/(Si)$_m$ \ac{sls}. Mass periodicity in the growth direction of \ac{sls} has long been recognized at the origin of phenomena such as zone folding and quantum confinement \cite{Esaki1986AWells}. Therefore, in order to eliminate \ac{sl} symmetry as an origin of the $E$\textsubscript{4$\tau$}, it is crucial to address any other plausible mechanisms. Firstly, quantum confinement should be excluded. Indeed, as \SiGe{}{} thickness increases (\textit{m} decreases), the $E$\textsubscript{4$\tau$} transition energy decreases at a rate smaller than a confinement\hyp{}related effect. For instance, consider the following hypothetical \ac{sl}, with \textit{m}$=\;3$, similar to $S_3$, with a fixed Si barrier thickness of $\SI{6}{\nano\meter}$ and a variable $t_w$ (between $5$ and $\SI{9}{\nano\meter}$). While suppressing any interfacial broadening, the absorption band edge redshifts within an energy range of $\SI{10}{\milli\electronvolt}$ as $t_w$ increases; see \textcolor{blue}{Supplemental Material \cite{Note1}}. The absence of any intrinsic additional peak below $\SI{3}{\electronvolt}$ proves that quantum confinement cannot be responsible of the observed $E$\textsubscript{4$\tau$} transition.
Secondly, the observed CP transition located between $\SI{2}{\electronvolt}$ and $\SI{2.5}{\electronvolt}$ can neither be explained by any of the $E_1$ \ac{cp} of the individual layers (Si or \SiGe{}{}, be it relaxed or strained) nor by a superposition of the $E_1+\Delta_1$ transition of \SiGe{}{} and the $E_1$ transition in Si. The $E_1$ transition in \SiGe{1-x}{x} was found to be located between $3.075$ and $\SI{2.850}{\electronvolt}$ for a Ge content between 20\% and 30\% which is still higher in energy than the observed $E$\textsubscript{4$\tau$} transition \cite{Bahng2001EvolutionAlloys}. Besides, the $E_1+\Delta_1$ transition in \SiGe{}{} alloys is only resolved for a Ge content above 32\%, which eliminates any possibility for the $E_1+\Delta_1$ transition to originate from within the \SiGe{}{} sublayer, as all \ac{sls} have an average Ge content below 30\% \cite{Ferrieu2000SpectroscopicIndustry,Lange1996DielectricSi001,Pickering1993SpectroscopicLayers,RajaMuthinti2012Effects001}. 
Thirdly, if the $E_1$ and $E_1+\Delta_1$ transition in Ge are considered, then a superposition with $E_1$ transition from Si or SiGe can lead to a reasonable explanation of the $E$\textsubscript{4$\tau$} transition. However, this hypothesis is unlikely due to the absence of any Ge segregation near the interfaces as confirmed by \ac{apt}.  
Moreover, similar to early studies on Si \cite{Lautenschlager1987TemperatureSilicon}, a $2D$ \ac{cp} was used for all \ac{cp}s except the $E$\textsubscript{4$\tau$} transition, for which a $3D$ \ac{cp} was used. This observation is confirmed throughout the analysis of all \ac{sls}, where a $3D$ transition gave a good quality fit with $R^2$ higher than $0.985$. The dashed-gray lines in Fig. \ref{fig:fig3d2e2} show the shift in the critical energy peak position as  the \ac{sl} periodicity changes. Increasing the periodicity up to $12$ periods leads to a small blueshift of $\SI{7.6}{\milli\electronvolt}$ $E$\textsubscript{4$\tau$}, followed by a redshift at the highest periodicity $\textit{m}=\;16$. This anomalous \ac{cp} is linked to the higher growth temperature of $S_{16}$ as compared to that of $S_3$. Higher growth temperatures induce a larger interfacial width \cite{Mukherjee20203DSuperlattices}. Note that Schmid \textit{et al.} \cite{Schmid1992OpticalSuperlattices} showed a transition near $\SI{2.49}{\electronvolt}$ for a $\SI{1}{\micro\meter}$ Ge$_7/$Si$_3$ \ac{sl}, which was attributed to the $E_1$ transition in the Ge-rich alloy. Note that in this work, the studied \ac{sls} are much thinner ($\sim \SI{60}{\nano\meter}$) with a Ge content below 30\%, thus excluding the possibility that the $E$\textsubscript{4$\tau$} emanates from Si or \SiGe{1-x}{x} individual layers. Therefore, the $E$\textsubscript{4$\tau$} transition can only be an interface-related transition. Indeed, a further confirmation of this assessment is given by the absorption coefficients of $S_3$ and $S_{16}$, shown in panel \ref{fig:fig1theory}(c, d), evaluated using experimental parameters (see caption of Fig. \ref{fig:fig1theory}). This agreement between theory and experiment is further confirmed by examining the behavior of annealed \ac{sls}, as discussed below.    
\par
\begin{figure}[t]
    \centering
    \includegraphics[width=0.9\columnwidth]{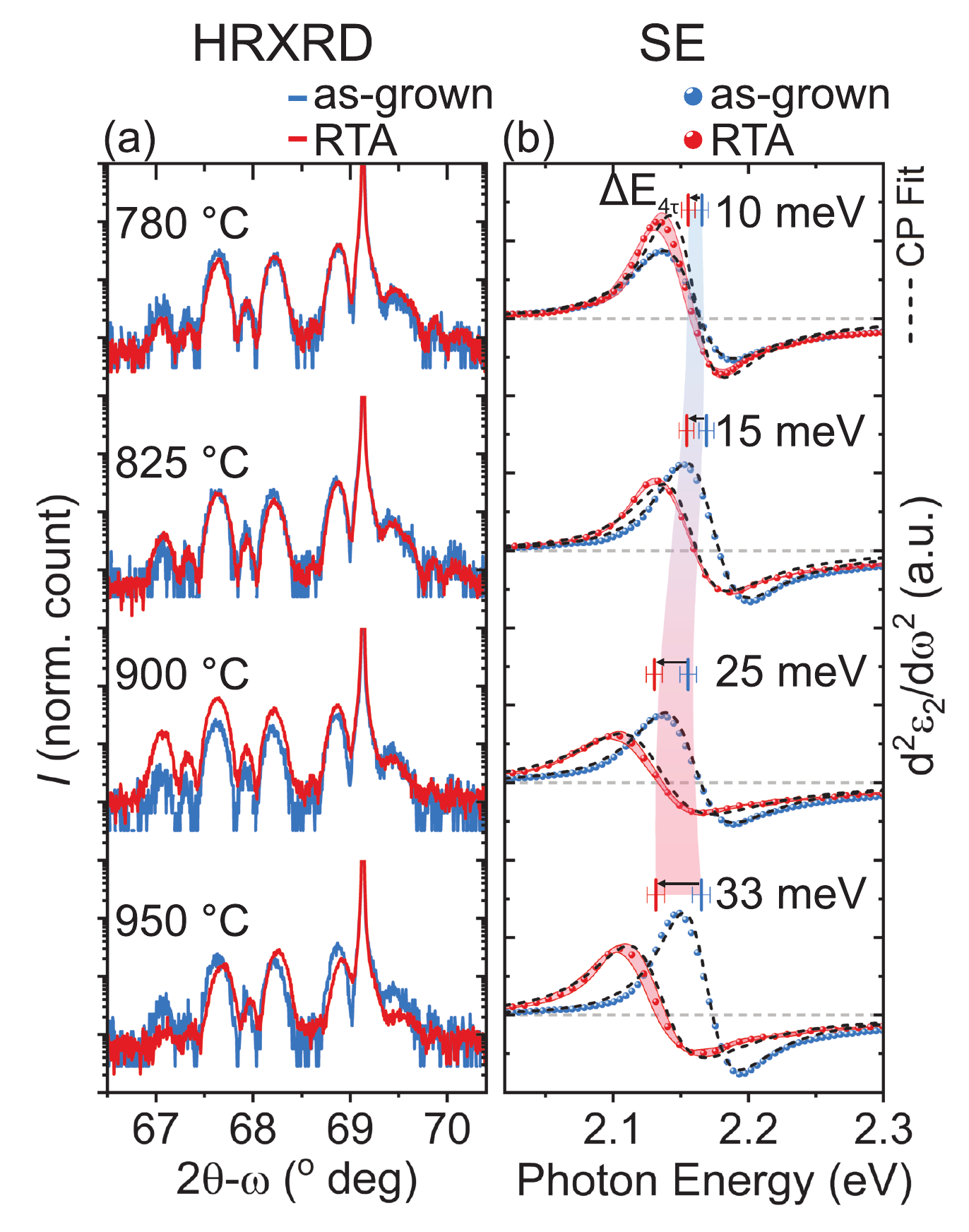}
    \caption{\textbf{Properties of annealed \ac{sls}.} (a) XRD $\omega-2\theta$ scans around the $(004)$ diffraction order of the as-grown (blue) and annealed (red) $S_3$ \ac{sl}. The annealing temperature was increased from $\SI{780}{\degree}$C to $\SI{950}{\degree}$C. (b) Second derivative statistical analysis of $\varepsilon_2$ is presented through the standard deviation $(4\sigma)$ (red error band). The standard deviation was also used during the $E$\textsubscript{4$\tau$} \ac{cp} lineshape fit (dashed lines) for both as-grown (blue) and annealed (red) $S_3$ samples. The interfacial broadening \ac{cp} energy shift $\Delta E$\textsubscript{4$\tau$} was analyzed as a function of the annealing temperature. The graded colored band (light blue to red) visualizes the redshift of the $E$\textsubscript{4$\tau$} \ac{cp} as the annealing temperature increases confirming the sensitivity of this peak to interfacial broadening.}
    \label{fig:fig4rta}
\end{figure}

 To enhance the interfacial broadening, \ac{sls} were subjected to rapid thermal annealing in the $\SI{780}{\degree}$C to $\SI{950}{\degree}$C temperature range under a flowing $N_2$ ambient gas for $50$ seconds. Fig. \ref{fig:fig4rta}(a) displays the $2\theta-\omega$ \ac{hrxrd} scans around the $(004)$ diffraction order of the as-grown (blue) and annealed (red) $S_3$ \ac{sl}. The presence of small intensity thickness fringes between the \ac{sl} satellite peaks, with an angular spacing inversely proportional to the total \ac{sl} thickness, is characteristic of pseudomorphic stacks with abrupt interfaces \cite{Py2011CharacterizationAnnealing}. As the annealing temperature increases, the well defined \ac{sl} peaks remain observable at the same angular position as for the as\hyp{}grown \ac{sl}. Besides, the thickness fringes become slightly less clear, which is an indication of a small interdiffusion. At the highest temperature ($\SI{950}{\degree}$C), the \ac{sl} peaks tend to become weaker and shift slightly toward the Si peak due to a larger interdiffusion. The corresponding $d^2\varepsilon_2/d\omega^2$ around the $E$\textsubscript{4$\tau$} \ac{cp} transition are exhibited in Fig. \ref{fig:fig4rta}(b).The vertical blue and red lines represent the $E$\textsubscript{4$\tau$} \ac{cp} peak position for the as\hyp{}grown and annealed \ac{sl}, respectively. Interestingly, the transition energy shift $\Delta E$\textsubscript{4$\tau$}$( =E_{4\tau}^{{\SI{300}{\degree}C}}-E_{4\tau}^{T})$ increases as a function of the annealing temperature from $\SI{10}{\milli\electronvolt}$ at $\SI{780}{\degree}$C to $\SI{33}{\milli\electronvolt}$ at $\SI{950}{\degree}$C. This redshift agrees well with the predicted theoretical results shown in Fig. \ref{fig:fig1theory}(c). Besides, the $E$\textsubscript{4$\tau$} \ac{cp} broadening $(\Gamma$ in $meV)$ increases with temperature from $\SI{30}{\milli\electronvolt}$ to $\SI{50}{\milli\electronvolt}$ which in qualitative agreement with the expected temperature effect on \ac{cp} \cite{Lautenschlager1987TemperatureSilicon}. It is also worth noting that \ac{hrxrd} seems to be less sensitive to interfacial broadening upon annealing at low temperature, whereas the detection of the optical fingerprint is rather straightforward hinting to the possibility to exploit the observed redshift in the $E$\textsubscript{4$\tau$} \ac{cp} as a sensitive interface metrology. Indeed, combining the theoretical framework (Fig. \ref{fig:fig1theory}c) and the measured shift $\Delta E$\textsubscript{4$\tau$} (Fig. \ref{fig:fig4rta}b) yields a straightforward and non-destructive method to extract the interfacial broadening width in the annealed samples. A logistic regression between $\Delta E$\textsubscript{4$\tau$} and (4$\tau$) was established based on the 14\hyp{}band $k\cdot p$ where $\Delta E$\textsubscript{4$\tau$} $(meV)=73.2/(1+e^{-15.9((4\tau(nm))-0.8)})$. This demonstrates an increase in the interfacial broadening from $\SI{0.71}{\nano\meter}$ to $\SI{0.81}{\nano\meter}$ as the annealing temperature increases from $\SI{780}{\degree}$C to $\SI{950}{\degree}$C. See the \textcolor{blue}{Supplemental Material \cite{Note1}} for a detailed analysis.
\UseRawInputEncoding
\par In summary, atomic-level interfacial broadening was found to create localized energy states in heterostructures. This phenomenon was predicted theoretically through a  rigorous theoretical formalism providing a qualitative and quantitative description of the absorption coefficient $\alpha$ of \ac{sls} in presence of smeared-out interfaces. The experimental measurements of \ac{cp} provided a direct evidence of this behavior and identified the associated optical transitions between $2$ and $\SI{2.5}{\electronvolt}$. Furthermore, thermal annealing-induced interfacial broadening confirmed  that these transitions are interface\hyp{}induced. This optical fingerprint lays the foundation for a sensitive, non-destructive probe of the atomic-level broadening of interfaces.\\ 

\bigskip
\noindent {\textbf{ACKNOWLEDGEMENTS}}.
The authors thank Mahmoud Atalla and Sebastien Koelling for fruitful discussions, Bill Baloukas for help with the spectroscopic ellipsometry measurements, J\'er\^ome Nicolas for help with the HRXRD measurements. O.M. acknowledges support from NSERC Canada (Discovery, SPG, and CRD Grants), Canada Research Chairs, Canada Foundation for Innovation, Mitacs, and PRIMA Qu\'ebec.\\

\medskip
\noindent {\textbf{AUTHORS INFORMATION}}.
Correspondence and requests for materials should be addressed to~O. Moutanabbir.

\medskip
\noindent {\textbf{AUTHOR CONTRIBUTIONS}}.
A.A. carried out the optical characterization. G.F. developed the $14-$band $k\cdot p$ formalism. S.M. performed the APT studies. M.B. carried out the epitaxial growth of the SLs. O.M. led this research. All authors commented on the manuscript.\\

\medskip
\noindent {\textbf{DATA AVAILABILITY}}.
The data that support the findings of this study are available from the corresponding authors upon reasonable request.\\

\bigskip

\bibliography{references_main.bib, Reference_SI.bib} 
\bibliographystyle{apsrev4-2} 

\end{document}